\documentclass[prX,showpacs,nofootinbib]{revtex4}
\newcommand{\be}{\begin{equation}}
\newcommand{\ee}{\end{equation}}
\newcommand{\ba}{\begin{eqnarray}}
\newcommand{\ea}{\end{eqnarray}}

\def\a{\alpha}
\def\b{\beta}
\def\d{\delta}
\def\e{\epsilon}

\def\g{\gamma}
\def\h{\eta}
\def\j{\psi}

\def\m{\mu}
\def\n{\nu}

\def\p{\pi}

\def\r{\rho}

\def\P{\Pi}

\def\S{\Sigma}

\newcommand{\ov}{\overline}

\newcommand{\aand}{\;\;\;\mbox{and}\;\;\;}
\newcommand{\pa}{\partial}
\newcommand{\sx}{\sigma_x}
\newcommand{\sy}{\sigma_y}
\newcommand{\sz}{\sigma_z}
\def\sl#1{\rlap{\hbox{$\mskip 1 mu /$}}#1}
\def\Sl#1{\rlap{\hbox{$\mskip 3 mu /$}}#1}
\def\I{\leavevmode\hbox{\small1\kern-3.8pt\normalsize1}}

\begin{document}

\title{No parity anomaly in massless QED$_3$: \\
A BPHZL approach}
\author{Oswaldo M. Del Cima}
\email{wadodelcima@if.uff.br}
\affiliation{Universidade Federal Fluminense (UFF),\\
P\'olo Universit\'ario de Rio das Ostras (PURO),\\
Departamento de Ci\^encia e Tecnologia,
Rua Recife s/n - 28890-000 - Rio das Ostras - RJ - Brazil.}

\author{Daniel H.T. Franco}
\email{dhtfranco@gmail.com}
\affiliation{Universidade Federal de Vi\c cosa (UFV),\\
Departamento de F\'\i sica - Campus Universit\'ario,\\
Avenida Peter Henry Rolfs s/n - 36570-000 -
Vi\c cosa - MG - Brazil.}

\author{Olivier Piguet}
\email{opiguet@pq.cnpq.br} 
\affiliation{Universidade Federal do Esp\'\i rito Santo (UFES),\\
CCE, Departamento de F\'\i sica,\\
Campus Universit\'ario de Goiabeiras - 29060-900 - Vit\'oria - ES - Brazil.}

\author{Manfred Schweda}
\email{mschweda@tph.tuwien.ac.at}  
\affiliation{Institut f\"ur Theoretische Physik,\\
Technische Universit\"at Wien (TU-Wien),\\
Wiedner Hauptstra{\ss}e 8-10 - A-1040 - Vienna - Austria.}

\date{\today}

%===================================================================
\begin{abstract}
In this letter we call into question the perturbatively parity breakdown 
at $1$-loop for the massless QED$_3$ 
frequently claimed in the literature.
As long as
perturbative quantum field theory is concerned, whether a parity
anomaly owing to radiative corrections exists or not will be
definitely proved by using a renormalization method independent of
any regularization scheme. Such a problem has been investigated in the 
framework of BPHZL renormalization method, by adopting the Lowenstein-Zimmermann 
subtraction sche\-me. The
$1$-loop parity-odd contribution to the vacuum-polarization tensor
is explicitly computed in the framework of the BPHZL
renormalization method. It is shown that a Chern-Simons term is
generated at that order induced through the infrared subtractions -- 
which violate parity. We show then that,  
what is called ``parity anomaly", is in
fact a parity-odd counterterm needed for restauring parity.
\end{abstract}
\pacs{11.10.Gh, 11.15.-q, 11.15.Bt, 11.15.Ex}
\maketitle

%===================================================================
%%%%%%%%%%%%%%%%%%%%%%%%%%%%%%%%%%%%%%%%%%%%%%
\section{Warm up: the massless QED$_3$ at 1-loop}
%%%%%%%%%%%%%%%%%%%%%%%%%%%%%%%%%%%%%%%%%%%%%%

It has been frequently addressed in the literature, since the work
of Ref.\cite{redlich}, that even perturbatively, parity is broken
in massless QED$_3$, such a breaking being achieved by a Chern-Simons
term that appears in the 1-loop contribution to the vacuum-polarization
tensor ($\P_{1\rm reg}^{\m\n}$) through the use of the Pauli-Villars (PV) 
regularization scheme:
\be
\P_{1\rm reg}^{\m\n}(p)=-{\frac{e^2}{16}}~{\frac{\h^{\m\n}p^2 - p^\m
p^\n}{\sqrt{p^2}}}+i\frac{e^2}{2\p}~\e^{\m\n\r}p_\r{\lim_{M\rightarrow\infty}} 
\frac{M}{|p|}\arcsin{\frac{|p|}{\sqrt{p^2-4M^2}}}
~.
\ee

The only superficially divergent 1-loop graphs present in massless
QED$_3$ are the vacuum-polarization tensor ($\P_1^{\m\n}$) and the
massless fermion self-energy ($\S_1$), with their degree of divergence
given by $\d(\P_1)=1$ and $\d(\S_1)=0$, respectively. In spite of their
na\"\i ve degree of divergence, they are in fact finite, the 1-loop Feynman
graphs give rise only to Speer-type integrals which do not develop
poles in the analytic continuation from $d\rightarrow 3$
\cite{speerint,leibbrandt,delbourgo}.

The 1-loop Feynman graphs for the vacuum-polarization (in the
Feynman gauge) and the fermion self-energy are given by: 
\ba
\P_1^{\m\n}(p)&=&-{\rm Tr}\int{\frac{d^3k}{(2\p)^3}}ie\g^\m
\left[i{\frac{\sl k}{k^2}}\right] ie\g^\n \left[i{\frac{\sl k -
\sl p}{(k-p)^2}}\right],\\
\S_1(p)&=&\int{\frac{d^3k}{(2\p)^3}}ie\g^\m\left[i{\frac{\sl
k}{k^2}}\right] ie\g^\n
\left[-i{\frac{\h_{\m\n}}{(k-p)^2}}\right]. 
\ea 
%By using the
%dimensional reguralization integrals~\cite{dr}, since only
%Speer-type integrals stem from those diagrams
%\cite{speerint,leibbrandt,delbourgo}, we get
%\be
%\P_1^{\m\n}(p)=-{\frac{e^2}{16}}~ {\frac{\h^{\m\n}p^2 - p^\m
%p^\n}{\sqrt{p^2}}} \aand \S_1(p)={\frac{e^2}{16}}~{\frac{\sl
%p}{\sqrt{p^2}}}~. \ee

At this point it should be stressed that neither gauge invariance
nor parity are broken by dimensional regularization (DR)~\cite{dr,delbourgo}:
\be
\P_1^{\m\n}(p)=-{\frac{e^2}{16}}~ {\frac{\h^{\m\n}p^2 - p^\m
p^\n}{\sqrt{p^2}}} \aand \S_1(p)={\frac{e^2}{16}}~{\frac{\sl
p}{\sqrt{p^2}}}~. 
\ee
Therefore, some claims
found in the literature that Pauli-Villars regularization must be
used because it does not break gauge invariance, contrary to the
dimensional regularization case, are not true in the light of perturbation
theory for massless QED$_3$. Moreover, Pauli-Villars breaks parity,
dimensional regularization does not, both preserve gauge invariance,
which of them is more suitable in the quantization of the massless
QED$_3$? Such a question makes no sense if both schemes are properly used. 
Fortuitous breakings
of gauge symmetry happens when there is no invariant regularization
scheme available, however, the Quantum Action Principle~\cite{qap,brs,pigsor}
guarantees that they can be completely eliminated, when gauge anomaly
is absent, by the introduction of noninvariant local counterterms
at each perturbative order.
The use of DR is quite well analized by Delbourgo and Waites for the 
three-dimensional QED$_3$~\cite{delbourgo}. Pimentel and Tomazelli ask in Ref.\cite{pimentel}, ``what's wrong with Pauli-Villars 
regularizarion in QED$_3$?'', there they present the correct use, which has to satisfy some necessary consistency conditions, of PV regularization in three space-time dimensions.  
The important issue of
``how superrenormalizable interactions cure their infrared divergences''
is analized by Jackiw and Templeton~\cite{jackiw}. As will be shown in a subsequent work~\cite{finitenessQED3}, 
through the use of the algebraic renormalization method~\cite{pigsor}, 
the massless QED$_3$ is in fact 
perturbatively finite\footnote{The authors of~\cite{delbourgo} have
already drawn attention to that and to the absence of a parity anomaly
as well, this was also pointed out by Rao and Yahalom~\cite{rao}.},
no regularization scheme is required at all. Furthermore,
an anomaly is not an ambiguity, it does not depend on which kind of
regularization scheme is made use of.

%What is called parity anomaly in the massless QED$_3$ is nothing
%more than a by-product of an inappropriate and inconsistency use of
%the Pauli-Villars regularization scheme, namely,   

%{\bf {\underline{1-loop $\P_1^{\m\n}$ using 
%different regularizations}}}
%\begin{enumerate}  
%   \item Dimensional Regularization {\small (Leibbrandt, 
%Del\-bourgo-Wai\-tes)}:
%{
%\[
%\P_{1\rm reg}^{\m\n}(p)=-{\frac{e^2}{16}}~{\frac{\h^{\m\n}p^2 - p^\m
%p^\n}{\sqrt{p^2}}}~,
%\] }
%   \item The { wrong} Pauli-Villars Regularization {\small (Red\-lich, 
%Du\-nne, many other followers)}:
%{
%\ba
%&&\P_{1\rm reg}^{\m\n}(p)=-{\frac{e^2}{16}}~{\frac{\h^{\m\n}p^2 - p^\m
%p^\n}{\sqrt{p^2}}}+\nonumber\\
%&&+~i\frac{e^2}{2\p}~\e^{\m\n\r}p_\r{\lim_{M\rightarrow\infty}} 
%\frac{M}{|p|}\arcsin{\frac{|p|}{\sqrt{p^2-4M^2}}}
%~,\nonumber
%\ea }
%   \item The { correct (consistent)} Pauli-Villars Regularization {\small (Pi\-men\-tel-Tomaze\-lli)}:
%{
%\[
%\P_{1\rm reg}^{\m\n}(p)=-{\frac{e^2}{16}}~{\frac{\h^{\m\n}p^2 - p^\m
%p^\n}{\sqrt{p^2}}}~,
%\] }
%\end{enumerate}

%%%%%%%%%%%%%%%%%%%%%%%%%%%%%%%%%%%%%%%%%%%%%%%%%%%
{\section{BPHZL: 1-loop vacuum polarization}}
%%%%%%%%%%%%%%%%%%%%%%%%%%%%%%%%%%%%%%%%%%%%%%%%%%%

In order to clarify the matter in an unambiguous way, we will perform
an explicit 1-loop computation in the Zimmermann's momentum space subtraction
scheme -- BPHZ~\cite{BPHZ}, which doe not use any regularization procedure.
Due to  to the presence of massless fields 
($\j$ and $A_\m$), the momentum subtraction scheme modified by
Lowenstein and Zimmermann -- BPHZL~\cite{lz} -- has to be adopted in order 
to deal with the infrared (IR) divergences 
that otherwise would arise in the process of ultraviolet (UV) subtractions.

The action for the massless QED$_3$ ($\S_{\rm QED_3}$), 
with the gauge invariant BPHZL mass terms added, is given by:
\be
\S^{(s-1)}=\int{d^3 x} \left\{
-\frac{1}{4}F^{\m\n}F_{\m\n}  + {\frac\m2}(s-1)\e^{\m\n\r}A_\m\pa_\n A_\r +
 i {\ov\j} {\Sl D}\j - m(s-1){\ov\j}\j - \frac{1}{2\a}(\pa_\m A^\m)^2\right\}~,
\label{action}
\ee 
where ${\Sl D}\j\equiv(\sl\pa+ie\Sl{A})\j$ and $e$ is a dimensionful
coupling constant. The BPHZL parameter $s$ lies in
the interval $0\le s\le1$ and plays the role of an additional
subtraction variable (as the external momentum) in the BPHZL
renormalization program. The massless theory is revored by putting
$s=1$ at the end of the calculations. At the classical level, the massless 
classical action for QED$_3$ is recovered for $s=1$: 
\be
\S_{\rm QED_3}=\S^{(s-1)}|_{s=1}\,.
\label{qed3}
\ee
Let us now establish some conventions and useful identities
necessary to compute the 1-loop parity-odd contribution to the vacuum-polarization tensor:
\ba
&&\h_{\m\n}={\rm diag}(+ - -)~,~\g^\m=(\sx,i\sy,-i\sz)~,\nonumber\\[3mm]
&&\{\g^\m,\g^\n\}=2\h^{\m\n}{\mathbf 1}~,~\g^\m\g^\n=\h^{\m\n}{\mathbf 1} + i\e^{\m\n\r}\g_\r~,\nonumber\\
&&\e^{\m\a\r}\e_{\n\b\r}=\d_\n^\m \d_\b^\a - \d_\b^\m \d_\n^\a \aand 
{\rm Tr}(\g^\m\g^\n\g^\r)=i\e_{\m\n\r}{\mathbf 1}~.
\ea 

For our 1-loop computations we need the fermion propagator, 
it reads
\be
\left\langle\ov\j(k)\j(k)\right\rangle=
i\frac{{\sl k}+m(s-1)}{k^2-m^2(s-1)^2}~.
\ee 
The 1-loop vacuum-polarization tensor, $\P_1^{\m\n}(p,s)$:
\ba
\P_1^{\m\n}(p,s)&=&\int{\frac{d^3k}{(2\p)^3}}
I_1^{\m\n}(p,k,s)\nonumber\\
&=&-e^2{\rm Tr}\int{\frac{d^3k}{(2\p)^3}}\g^\m
\frac{{\sl k}+m(s-1)}{k^2-m^2(s-1)^2} \g^\n\frac{{\sl k -
\sl p}+m(s-1)}{(k-p)^2-m^2(s-1)^2}~,
\ea 
has the following UV and IR subtraction degrees, $\d(\P)=1$ and $\r(\P)=1$, respectively.

In the BPHZL scheme the subtracted integrand, $R_1^{\m\n}(p,k,s)$, is written in terms of the unsubtracted one, $I_1^{\m\n}(p,k,s)$, as 
\ba
R_1^{\m\n}(p,k,s)&=&(1-t^{0}_{p,s-1})(1-t^{1}_{p,s})
I_1^{\m\n}(p,k,s)\nonumber\\
&=&(1-t^{0}_{p,s-1}-t^{1}_{p,s}+t^{0}_{p,s-1}t^{1}_{p,s})
I_1^{\m\n}(p,k,s)~,
\ea  
where $t^{d}_{x,y}$ is the Taylor series about $x=y=0$ 
to order $d$ if $d\geq 0$.

Thus, for our purposes, by assuming $s=1$, the subtracted integrand, $R_1^{\m\n}(p,k,s)$, reads
\be
R_1^{\m\n}(p,k,1)=\underbrace{I_1^{\m\n}(p,k,1)}_{\rm parity-even}-\underbrace{I_1^{\m\n}(0,k,1)}_{\rm parity-even}-
\underbrace{p^\r\frac{\pa}{\pa p^\r}I_1^{\m\n}(0,k,0)}_{\rm parity-odd~terms}~,
\ee 
where 
\ba
&&I_1^{\m\n}(p,k,s)=
-e^2{\rm Tr}~\g^\m
\frac{{\sl k}+m(s-1)}{k^2-m^2(s-1)^2} \g^\n\frac{{\sl k -
\sl p}+m(s-1)}{(k-p)^2-m^2(s-1)^2}~,\\
&&I_1^{\m\n}(p,k,1)=-e^2{\rm Tr}~\g^\m
\frac{{\sl k}}{k^2} \g^\n\frac{{\sl k -
\sl p}}{(k-p)^2}~,\\[3mm]
&&I_1^{\m\n}(0,k,1)=-e^2{\rm Tr}~\g^\m
\frac{{\sl k}}{k^2} \g^\n\frac{{\sl k}}{k^2}~,\\
&&p^\r\frac{\pa}{\pa p^\r}I_1^{\m\n}(0,k,0)=-e^2{\rm Tr}~\g^\m\frac{{\sl k}-m}{k^2-m^2} \g^\n
\left[-\frac{\sl p}{k^2-m^2}+2p\cdot k 
\frac{{\sl k}-m}{(k^2-m^2)^2} \right]
~.
\ea 

Bearing in mind that in the intermediary steps of the BPHZL 
subtraction scheme parity is explicitly broken, when $s=0$ is assumed, a Chern-Simons term could be expected as a parity-odd 1-loop counter-term due to the parity-noninvariant nature of the BPHZL renormalization scheme. In fact, such a CS term, generated at the 1-loop correction to the vacuum-polarization tensor, arises in the process of IR subtractions performed so as to subtract IR divergences induced by the UV subtractions.

The parity-odd contribution, $R_{1\rm odd}^{\m\n}(p,k,1)$, stemming from 
the subtracted integrand, is given by
\be
R_{1\rm odd}^{\m\n}(p,k,1)=
e^2 m~\e^{\m\n\r}p_\r\frac{1}{(k^2-m^2)^2}\,.
\ee
Then, the parity-odd 1-loop subtracted (finite) vacuum-polarization tensor, 
$\P_{1\rm odd}^{\m\n\rm (sub)}(p,1)$, reads
\ba
\P_{1\rm odd}^{\m\n\rm (sub)}(p,1)&=&
e^2 m~\e^{\m\n\r}p_\r\int{\frac{d^3k}{(2\p)^3}}
\frac{1}{(k^2-m^2)^2}\nonumber\\
&=&i\frac{e^2}{8\p}\frac{m}{|m|}~\e^{\m\n\r}p_\r~.
\label{odd-term}\ea 
The appearence of this  contribution may however be
compensated by adding to the action a gauge invariant
local Chern-Simons counterterm:
\be
\frac{e^2}{8\p}\frac{m}{|m|}\int{d^3 x}\, \e^{\m\n\r}A_\m\pa_\n A_\r~\,.
\ee
%So, the 1-loop effective action, subtracted via BPHZL, can be written as
%\be
%\G^{(s-1)}_{1\rm eff}\bigg|^{s=1}=\S_{\rm QED_3}+\G_{1\rm even}+
%\int{d^3 x}~\bigg[\a - \frac{e^2}{8\p}\frac{m}{|m|}\bigg]\e^{\m\n\r}A_\m\pa_\n A_\r~,
%\ee 
%such that, together with the 2-point normalization condition for the 
%massless gauge field $A_\m$:
%\be
%\frac{\pa}{\pa p^2}
%\G^{(s-1)}_{A^TA^T}(p^2)\bigg|^{s=1}_{p^2=0}=1~,
%\ee
%the coefficient $\a$ of the CS counterterm is fixed as follows:
%\be
%\a = \frac{e^2}{8\p}\frac{m}{|m|}~.
%\ee
Therefore, there is no parity-odd 1-loop contribution to the vacuum-polarization 
tensor $\P_{1}^{\m\n\rm (sub)}(p,1)$.
%$\P_{1\rm odd}^{\m\n\rm (sub)}(p,1)$, is a counter-term. 
\\
\\
%%%%%%%%%%%%%%%%%%%%%%%%%%%%%%
\section{Conclusion}
%%%%%%%%%%%%%%%%%%%%%%%%%%%%%%
In the BPHZL renormalization scheme used in the present work, the odd term 
(\ref{odd-term}) appears as a consequence of the IR subtractions, 
performed at $s=0$, which explicitly break 
parity. However, as we have seen, we can remove it through the 
introduction of a Chern-Simons local counterm and thus recover parity
symmetry. This is more an illustration of the restablishing of a
symmetry which has been broken by the subtraction or regularization
procedure used. This is well known for the case of continuous 
symmetry~\cite{brs,pigsor}. To the best of our knowledge, \textit{this is new for a
discrete symmetry}.

In short, such a Chern-Simons term could never be interpreted as a
``parity anomaly''.
We thus conclude that no parity anomaly exists perturbatively in massless
QED$_3$.

%%%%%%%%%%%%%%%%%%%%%%%%%%%%%%%%%%
\subsection*{Acknowledgements}
%%%%%%%%%%%%%%%%%%%%%%%%%%%%%%%%%%
O.M.D.C. dedicates this work to his kids, Vittoria and Enzo, to his mother, Victoria, and to Clarisse Czertok. 
D.H.T.F. dedicates this work to Prof. Lincoln Almir Amarante Ribeiro (in memorium).

% it is a c%ounter-term remnant of the BPHZL subtraction scheme
%$\clubsuit$In fact, the study of the stability of action requires 
%that a Chern-Simons-like term
%already is present in action (\ref{action}) since the beginning.$\clubsuit$ 
%ounter-term remnant of the BPHZL subtraction scheme.batively in massless QED$_3$. 
%The reabsorption of the Chern-Simons term as a counter-term at 1-loop is 
%performed by fixing a coefficient ($\a$) through the 2-point normalization 
%condition for the massless gauge field $A_\m$, which is automatically 
%satisfied by the Lowenstein-Zimmermann subtraction scheme. 

% ----------------------------------------------------------------
\end{document}